\documentclass[onecolumn,aps,showpacs,showkeys,preprintnumbers,amsmath,amssymb,nofootinbib,superscriptaddress]{revtex4}
\usepackage{epsfig}
\usepackage{amsfonts}
\usepackage{dcolumn}

\usepackage{amsmath}

\def\beq{\begin{equation}}
\def\eeq{\end{equation}}
\def\bea{\begin{eqnarray}}
\def\eea{\end{eqnarray}}
\def\bi{\begin{itemize}}
\def\ei{\end{itemize}}

\let\al=\alpha
\let\be=\beta
\let\ga=\gamma

\let\de=\delta

\let\ep=\epsilon

\let\si=\sigma

\let\ta=\tau
\def\ph{\varphi}

\begin{document}
\author{Anders Basb\o ll}
\email{ab421@sussex.ac.uk} \affiliation{Department of Physics \& Astronomy, University of Sussex, Brighton, BN1 9QH, United Kingdom}
\title{A complete and minimal catalogue of MSSM gauge invariant monomials}
\date{October 2nd, 2009}

\begin{abstract}
We present a complete and minimal catalogue of MSSM gauge invariant monomials. That is, the catalogue of Gherghetta, Kolda and Martin is elaborated to include generational structure for all monomials. Any gauge invariant operator can be built as a linear combination of elements of the catalogue lifted to nonnegative integer powers. And the removal of any one of the monomials would deprive the catalogue of this feature. It contains 712 monomials - plus 3 generations of righthanded neutrinos if one extends the model to the $\nu$MSSM. We note that $\nu$MSSM flat directions can all be lifted by the 6th order superpotential - compared to the 9th order needed in MSSM.
\end{abstract}
\maketitle \pagebreak

\section{Introduction}
The scalar potential of the Minimal Supersymmetric Standard Model
(MSSM) consists of F-terms ($\sum_{\phi}|\partial W/\partial \phi|^2$) and D-terms ($\sum_{a}g_a^2/2|\sum_{\phi}\phi^\dagger T^a \phi|^2$) where $T^a, W$ are the gauge generators and the superpotential respectively - and $\phi$ are the scalar fields. The potential possesses a large number of D-flat directions \cite{Gherghetta:1995dv,Enqvist:2003gh}. Some of these are also F-flat when one considers the renormalisable superpotential only. If one allows higher order nonrenormalisable terms (only accepting R-parity as conserved) all these flat directions are lifted by different terms of different order. A catalogue of D-flat directions was presented by Gherghetta, Kolda and Martin in \cite{Gherghetta:1995dv} in which it was also shown at which order of possible nonrenormalisable terms in the superpotential each flat direction could be lifted (with no righthanded neutrinos). This catalogue of monomials is here expanded to include generational structure for all gauge invariant monomials (corresponding to D-flatness) - which is not as trivial as one might expect. F-flatness will only briefly be mentioned here. It will be noted that allowing the MSSM to be augmented by righthanded neutrinos ($\nu$MSSM) all flat directions will be lifted by terms in the 6th order of the superpotential - compared to the 9th order needed in the MSSM.

 The cosmological role of flat directions have been studied intensively. For instance the possibility of the flat directions creating the baryon assymmetry of the universe is investigated in
\cite{Affleck:1984fy,Linde:1985gh,Dine:1995uk} while the cosmological importance of flat direction vacuum
expectation values (VEVs) as a possible delayer of thermalisation has been investigated in 
 \cite{Allahverdi:2006iq,Allahverdi:2006wh,Olive:2006uw,Allahverdi:2006xh,Allahverdi:2006gralep,us,me,Allahverdi:2008,GuiMetRioRiva,Olive08,warsaw,xflat,bjorn,RioRiva,brand,dufaux,KasKaw,ShoeKuse,Gumruk}.


\section{Flat directions}\label{sNppp}

As pointed out in \cite{Gherghetta:1995dv} the D-flat space of the MSSM (no right-handed neutrinos) has 37 complex dimensions. This arises as 49 complex fields minus 12 real non-flat directions and 12 real gauge choices (1 of each for each gauge generator). 

In fact, the catalogue of \cite{Gherghetta:1995dv} is a catalogue over $SU(3)_c\times SU(2)_L\times U(1)_Y$ invariant monomials - of which there are many more than complex dimensionalities of the D-flat space. It is complete in the sense that including all possible generational structure to the monomials (which is largely omitted) any gauge invariant operator can be built as a linear combination of elements of the catalogue lifted to nonnegative integer powers. It is minimal in the sense that if removing any monomial \emph{with all  its generational structure} the completeness will be lost. It contains 28 types of monomials. The completeness of these are taken as a starting point for this work. The minimalness is confirmed here, since all types of monomials will be shown to contribute with degrees of freedom. The relation between flat directions and gauge invariance (sometimes mentioned as if the same) is best illustrated in an example. $H_u$ (The Higgs doublet with positive hypercharge) and $H_d$ (The Higgs doublet with negative hypercharge) can combine to make the gauge invariant product $H_u^\al H_d^\be \ep_{\al \be}=H_u^+H_d^--H_u^0H_d^0$. One can then assign VEV's to any of the terms (here there are only two, but in other cases, typically involving color, there are more) i.e. $|\langle H_u^+\rangle|=|\langle H_d^-\rangle|=\varphi, |\langle H_u^0\rangle|=|\langle H_d^0\rangle|=0$ or vice versa, and all the D-terms will be zero (because the fields with VEVs have equal magnitude, opposite sign hypercharge and opposite $SU(2)_L$-charge. As long as only one term is involved, flatness is independent of phase. If more terms are involved, the phases do matter. In fact, in the case above, the monomial breaks $SU(2)_L\times U(1)_Y$ to a new symmetry $U(1)_{new}$. This means four gauge symmetries are reduced to one and thus there are 3 independent non trivial D-terms. Demanding all the D-terms to be zero (flatness) leaves 5 degrees of freedom $H_u= \left(\begin{array}{c} \varphi_1 e^{i\si_1} \\ \varphi_2 e^{i\si_2}
\end{array} \right)$ and $H_d= \left(\begin{array}{c} \varphi_2 e^{i\si_3} \\ \varphi_1 e^{i(\si_3+\si_2-\si_1+\pi)}\end{array} \right)$ and after that you are left with 3 gauge choices - setting $H^+=\varphi_1 e^{i\si_1}$ to zero makes things much simpler, and $\si_2=\si_3$ is a nice choice, since in the gauge invariant product only the sum of the 2 phases matter. So we can use $H_u= \left(\begin{array}{c}0 \\ \varphi e^{i\si}
\end{array} \right)$ and $H_d= \left(\begin{array}{c} \varphi e^{i\si}\\ 0\end{array} \right)$. 
However, it is not always so simple - we showed in \cite{me} that in $(QQQ)_4LLLE$ ($Q$: lefthanded squarks [sloppy for scalar partner of lefthanded quark - this should cause no confusion], $L$: lefthanded sleptons, $E$: righthanded hypercharged sleptons) one can only gauge some of the phase differences between the participating VEV fields away. On the other hand, in $LLE=L_{i+1}^{\al}L_{i+2}^{\be}E_j\ep_{\al \be},[1\leq i,j\leq 3]$ there are clearly 9 monomials (3 choices of which $E$ is involved and 3 choices of which $L$ is NOT involved). However, $L$ and $E$ combined contain 9 complex dimensions - and the monomials break $SU(2)_L\times U(1)_Y$ completely. Thus we have 4 broken generators corresponding to 4 complex degrees of freedom being removed - by non-flatness (4 real d.o.f.) and gauge choices (4 real d.o.f.). This means that $L,E$ has D-flat complex dimentionality of 5, eventhough it is spanned by a set of 9 monomials. There are no linear combinations of the 9 that give zero (except for the trivial one). So this illustrates that there are many more monomials than complex dimensionality. This comes from the nonlinearity of the equations. Since all D-flatness comes from terms with field dimension 4 (that is all of the same dimension), any combination of fields that has zero potential can be scaled by the same factor. Thus the monomials $L_1L_2E_1$ and $L_1L_2E_2$ suggest the flatness of $|\langle\nu_e\rangle|=|\langle\mu\rangle|=\langle e^c\rangle|=a$ and $|\langle \nu_e\rangle|=|\langle \mu\rangle|=|\langle \mu^c\rangle|=b$ but linear combinations of these are not flat $|\langle \nu_e\rangle|=|\langle \mu\rangle|=(a+b),|\langle e^c\rangle|=a,|\langle \mu^c\rangle|=b$. So the flat directions suggested by the monomial are lines in field space - but 2 lines with zero potential do not (neccessarily) span a flat plane. This example illustrates the point of linearity, but not the difference in number of monomials and dimensionality. For $L_1,L_2,E_1,E_2$ contain 6 complex degrees of freedom (c.d.o.f.) and break all 4 electroweak generators -- leaving 2 c.d.o.f. -- equal to the number of monomials. In fact, setting $|\langle\nu_e\rangle|=|\langle\mu\rangle|=\sqrt{a^2+b^2}$ would give flatness for any values of $a,b$\footnote{phases have been omitted - but there are only one phase if one monomial, and there are two free phases in the just mentioned case.}. But augmenting the system by $L_3$ adds 2 c.d.o.f., no new broken symmetries and 4 new monomials. So we have 4.c.d.o.f. and 6 monomials. Imposing flatness and making gauge choices (such that it is easy to see there are 4 complex parameters) this system is \begin{eqnarray}&&\nonumber \nu_e=\ph_1e^{i\si_1},\nu_\mu=\ph_2e^{i\si_2},e^c=\ph_3e^{i\si_3},\mu^c=\ph_4e^{i\si_4},\ta =0,\\
&&e=\frac{\sqrt{\ph_3^2 +\ph_4^2}\ph_2}{\sqrt{\ph_1^2 + \ph_2^2}}e^{i\si_1},\mu=\frac{\sqrt{\ph_3^2 +\ph_4^2}\ph_1}{\sqrt{\ph_1^2 + \ph_2^2}}e^{i(\si_2+\pi)},\nu_\ta=\sqrt{\ph_3^2 +\ph_4^2-\ph_1^2-\ph_2^2}e^{i\si_1}
\end{eqnarray}
where the condition necessary for the last squareroot to be real should be imposed. We can, of course, refind monomials: $L_1L_2E_1$ corresponds to $\ph_1=\ph_3=1$, but the parameterisation is pretty bad for others -- $L_2L_3E_1$ corresponds to $\ph_3=1,\ph_1=a,\ph_2=a^2$ for $a\rightarrow 0$ and  $L_3L_1E_1$ corresponds to $\ph_3=1,\ph_1=a^2,\ph_2=a$ for $a\rightarrow 0$. The remaining 3 monomials come from interchange of $\ph_3$ and $\ph_4$. It is probably possible to make parameterisation where all monomials can be found without using limits, but the above parameterisation was chosen because it had the feature of having 4 fields being equal to one free complex parameter each.
\section{The monomials}
Here follows a list of the monomials. Color indices are $a,b...$, anticolor $\overline{a},\overline{b}...$. Family indices are latin letters $i,j,k...$, $SU(2)_L$ are greek letters $\al ,\be, \ga...$, ($\de , \ep$ are reserved for the Kronecker delta and maximally antisymmetric tensors) and other latin letters are used occassionally to mark other dimensionalities. Summation over repeated indices is implied if and only if one is lowered and the other is raised. The superfields and there scalars will be referred to with the same symbols - those not mentioned yet are righthanded squarks ($U$: negative hypercharge, $D$: positive hypercharge). We look at all $28$ monomial types from \cite{Gherghetta:1995dv} - but first the simplest ones: righthanded neutrinos.
\subsection{Righthanded neutrinos - $\nu$MSSM}
Since \cite{Gherghetta:1995dv} was published neutrinos have been shown to have mass. The most natural way to get that is by allowing righthanded neutrinos (which is actually mentioned as a possibility by \cite{Gherghetta:1995dv}). Righthanded neutrinos ($N$s) are gauge invariant monomials by themselves
\begin{equation}
(N)_i=N_i,\textbf{       }[1\leq i\leq 3]
\end{equation}
In fact, with these we will find the gauge invariant monomials of $\nu$MSSM.
\subsubsection{A note on F-flatness}
One interesting consequence of including righthanded neutrinos is that they get rid of the formally flattest directions of the MSSM. In \cite{Gherghetta:1995dv} it was shown that only two combinations of fields survive the superpotential of order 6. One is $D,L$ with monomials $DDDLL$ that is lifted by seventh order in $W$. The other is $Q,U,E$ with monomials $QQQQU, UUUEE, QUQUE$ - which is only lifted by order 9 in W. All these monomials have negative R-parity. But if multiplied by $N_i$ these products are in the 6th order in the superpotential - and these easily lift  the flat directions. It should be pretty obvious that no new very flat directions emerge. One can add $N$s to everything - but we always get new F-terms from the $N$s aswell. The major role of $N$ is, that any FD with negative R-parity will get a new positive R-parity partner from just adding $N$. This increase the number of terms that can break F-flatness dramatically. With any positive R-parity monomial, for each degree of freedom we get an F-term. So, in the $\nu$MSSM  every flat direction can be lifted by terms in the 6th order of the superpotential - in contrast to the 9th order needed in MSSM. Since the F-term comes from a squared field derivative, a term of order $n$ in the superpotential corresponds to a term of order $2n-2$ in the scalar potential itself. Thus, every flat direction should be lifted by 10th order in the potential instead of the 16th order.
\subsection{The trivial ones}
The easiest are the monomials where each superfield appears once or not at all. Here the dimensionality is just the product of the number of generations of the superfields. We have
\begin{eqnarray}
(LH_u)_i&=L_i^{\al}H_u^{\be}\ep_{\al \be},&[1\leq i\leq 3]\\
H_uH_d&=H_u^{\al}H_d^{\be}\ep_{\al \be}&\\
(QLD)_{i,j,k}&=Q_i^{\al ,a}L_j^{\be}D_k^{\overline{a}}\ep_{\al \be}\de_{a\overline{a}},&[1\leq i,j,k\leq 3]\\
(QH_uU)_{i,j}&=Q_i^{\al ,a}H_u^{\be}U_j^{\overline{a}}\ep_{\al \be}\de_{a\overline{a}},&[1\leq i,j\leq 3]\\
(QH_dD)_{i,j}&=Q_i^{\al ,a}H_d^{\be}D_j^{\overline{a}}\ep_{\al \be}\de_{a\overline{a}},&[1\leq i,j\leq 3]\\
(LH_dE)_{i,j}&=L_i^{\al}H_d^{\be}E_j\ep_{\al \be},&[1\leq i,j\leq 3]\\
(QLUE)_{i,j,k,l}&=Q_i^{\al ,a}L_j^{\be}U_k^{\overline{a}}E_l\ep_{\al \be}\de_{a\overline{a}},&[1\leq i,j,k,l\leq 3]\\
(QH_dUE)_{i,j,k}&=Q_i^{\al ,a}H_d^{\be}U_j^{\overline{a}}E_k\ep_{\al \be}\de_{a\overline{a}},&[1\leq i,j,k\leq 3]
\end{eqnarray}
\subsection{Simple antisymmetric ones}
Superfields that have exclusively $SU(3)_c$ or $SU(2)_L$ interactions must be antisymmetric when contracted with a superfield of the same kind. This means family indices must be different (and thus, since there are 3 generations, they can be named after the one not participating if the superfield participates twice, or needs no index at all if the superfield participates thrice). Below indices on Superfields are understood to be modulo 3 \footnote{or rather (index minus one) modulo 3, plus one}
\begin{eqnarray}
(LLE)_{i,j}&=L_{i+1}^{\al}L_{i+2}^{\be}E_j\ep_{\al \be},&[1\leq i,j\leq 3]\\
(UDD)_{i,j}&=U_i^{\overline{a}}D_{j+1}^{\overline{b}}D_{j+2}^{\overline{c}}\ep_{\overline{a}\overline{b}\overline{c}},&[1\leq i,j\leq 3]\\
(UUDE)_{i,j,k}&=U_{i+1}^{\overline{a}}U_{i+2}^{\overline{b}}D_{j}^{\overline{c}}E_k\ep_{\overline{a}\overline{b}\overline{c}},&[1\leq i,j,k\leq 3]\\
(DDDLL)_{i}&=D_1^{\overline{a}}D_{2}^{\overline{b}}D_{3}^{\overline{c}}L_{i+1}^{\al}L_{i+2}^{\be}\ep_{\overline{a}\overline{b}\overline{c}}\ep_{\al \be},&[1\leq i\leq 3]\\
(DDDLH_d)_{i}&=D_1^{\overline{a}}D_{2}^{\overline{b}}D_{3}^{\overline{c}}L_{i}^{\al}H_d^{\be}\ep_{\overline{a}\overline{b}\overline{c}}\ep_{\al \be},&[1\leq i\leq 3]
\end{eqnarray}
\subsection{$QUQUE$}
Clearly, the two $QU$s must be antisymmetric. But it leaves the question whether $Q_{i1}U_{j1}Q_{i2}U_{j2}E_k$ and $Q_{i1}U_{j2}Q_{i2}U_{j1}E_k$ (antisymmetric tensors omitted) are propotional. By inspection, they are not.
\begin{equation}
(QUQUE)_{i,j,k,l,m}=Q_{i}^{\al a}U_{j}^{\overline{a}}Q_{k}^{\be b}U_{l}^{\overline{b}}E_m\de_{a\overline{a}}
\de_{b\overline{b}}\ep_{\al \be},[1\leq i,j,k,l,m\leq 3,i<k \lor (i=k\land j<l)] 
\end{equation}
\subsection{$QUQD$}
Here the question is whether $Q_{i1}U_{j}Q_{i2}D_{k}$ and $Q_{i2}U_{j}Q_{i1}D_{k}$ (antisymmetric tensors omitted) are propotional. They are not.
\begin{equation}
(QUQD)_{i,j,k,l}=Q_{i}^{\al a}U_{j}^{\overline{a}}Q_{k}^{\be b}D_{l}^{\overline{b}}\de_{a\overline{a}}\de_{b\overline{b}}\ep_{\al \be},[1\leq i,j,k,l\leq 3]
\end{equation}
\subsection{The one with 2 $E$s}
$E$ is only charged under $U(1)_Y$ which is Abelian. Therefore $E_iE_j=E_jE_i$ and $UUUEE$ is six dimensional as pointed out in \cite{Gherghetta:1995dv}
\begin{equation}
(UUUEE)_{i,j}=U_1^{\overline{a}}U_{2}^{\overline{b}}U_{3}^{\overline{c}}E_iE_j\ep_{\overline{a}\overline{b}\overline{c}},[1\leq i\leq j\leq 3]
\end{equation}
\subsection{The ones with 4 superfields including 3 $Q$s}
When there are 4 Superfields of which 3 are $Q$s, obviously the $Q$s must contract color indices and 2 of them must contract $SU(2)_L$ indices. At first look this gives 27 choices of $Q$: 
$(QQQ)_{ijk}^{\al}=Q_{i}^{a \be}Q_{j}^{b \ga}Q_{k}^{a \al}\ep_{abc}\ep_{\be \ga}$. Clearly there is symmetry between the first two generation indices, but we will not use this yet. It is convenient to split the discussion in the cases of how many generations are present. First, 3 $Q$s from the same generation clearly gives zero due to the antisymmetric tensor for $SU(2)_L$. If there are 2 of one generation and one from a different one - let us call this the 2-1 case - it turns out  $QQQ_{iij}+QQQ_{iji}+QQQ_{jii}=0$ and $QQQ_{iji}-QQQ_{jii}=0$, whereas  $QQQ_{iij\textbf{h}}=\frac{QQQ_{iji}+QQQ_{jii}-2QQQ_{iij}}{\sqrt{6}}$ is free. While unneccessary for the discussion of the monomials, we choose to normalise such that antisymmetric tensors (or delta tensors) contracting $SU(3)_c$ and $SU(2)_L$ indices are normalised by one - whereas other linear combinations are normalised as if built of unitvectors in the superfield product space -- and we make basis vectors that are orthogonal to the basisvectors that span zero. That is, if $a,b,c$ are products of fields and $SU(3)_c,SU(2)_L$ antisymmetric tensors and $\overrightarrow{a}-\overrightarrow{b}=\overrightarrow{a}+\overrightarrow{b}+\overrightarrow{c}=0$ then  $\frac{\overrightarrow{a}+\overrightarrow{b}-2\overrightarrow{c}}{\sqrt{6}}$ is the correct normalisation.

Finally, there can be one $Q$ of each generation (1-1-1). There is symmetry between the first two indices\footnote{$SU(2)_L$ indices omitted}:  $QQQ_{ijk}-QQQ_{jik}=0$ leading us to define $QQQ_{123\textbf{i}}= (QQQ_{jki}+QQQ_{kji})/\sqrt{2}$ (j,k,i all different). Then $QQQ_{123\textbf{1}}+QQQ_{123\textbf{2}}+QQQ_{123\textbf{3}}=0$ and $QQQ_{123\textbf{7}}=(QQQ_{123\textbf{1}}-QQQ_{123\textbf{2}})/\sqrt{2}$ and $QQQ_{123\textbf{8}}=(QQQ_{123\textbf{1}}+QQQ_{123\textbf{2}}-2QQQ_{123\textbf{3}})/\sqrt{6}$ are free. Indices in bold (with values higher than three) do not represent a specific superfield (7,8 and $\textbf{e}$ in table below is used in order for these to look quite different from generational indices).
\begin{eqnarray}
(QQQL)_{i,i,j,k}&=QQQ_{iij\textbf{h}}^{\al}L_{k}^{\be}\ep_{\al \be},&[1\leq i,j,k\leq 3,i\neq j]\\
(QQQL)_{1,2,3,\textbf{e},k}&=QQQ_{123\textbf{e}}^{\al}L_{k}^{\be}\ep_{\al \be},&[1\leq k\leq 3,7\leq e\leq 8]\\
(QQQH_d)_{i,i,j}&=QQQ_{iij\textbf{h}}^{\al}H_d^{\be}\ep_{\al \be},&[1\leq i,j\leq 3,i\neq j]\\
(QQQH_d)_{1,2,3,\textbf{e}}&=QQQ_{123\textbf{e}}^{\al}H_d^{\be}\ep_{\al \be},&[7\leq e\leq 8]
\end{eqnarray}
\subsubsection{Are these monomials?}
Yes. At least in the same sense that \cite{Gherghetta:1995dv}'s $QQQ_\textbf{4}^{\al \be \ga}=Q_1^{\al a}Q_2^{\be b}Q_1^{\ga c}\ep_{abc}\ep^{ijk}$ is\footnote{multiplied by some other fields which are irrelevant to this discussion.}. The $\textbf{4}$ means it transforms as a $\textbf{4}$ under $SU(2)_L$\footnote{no contraction of $SU(2)$ indices among $Q$s} - whereas the product in the previous session transforms as a doublet ($\textbf{2}$). The generation antisymmetry is a completely different thing than the antisymmetry in the gauge couplings. In fact, the epsilon of generations is doing exactly the same as the linear combination $\frac{\overrightarrow{a}+\overrightarrow{b}-2\overrightarrow{c}}{\sqrt{6}}$ in the case before. Finding something that is orthogonal to those combinations that span zero. So the meaning of monomial is in fact just that you can build all gauge invariant expressions from linear combinations of the monomials raised to some nonnegative powers. Also, these definitions make it easier to define the monomials in the next section in a reasonable way.
\subsection{The one with 4 $Q$s}
$QQQQU$ is an interesting case. Ignoring the $U$ dimensionality of 3, one could think the $Q$ part is 24 dimensional - 8 from $(QQQ)_\textbf{2}$, as already seen, and 3 from the $Q$ from $QU$. Another guess could be to look at how many ways you can asign generation indices to 4 (unordered) $Q$s - leaving out the 4 identical $Q$s option - since the 3 from $QQQ_\textbf{2}$ cannot be identical. This gives 12 $Q$ possibilities. In fact, the correct answer is that the $Q$-part is 18 dimensional.

We can use the paramererisations from before - though we will not use the last 2 definitions (those including indices 7,8). Again, all 4 $Q$s cannot be the same generation. 3 $Q$s from the same generation has to be 2-1 from $QQQ$ and 1-0 from the $Q$ contracted with the $U$. That is $QQQ_{iij\textbf{h}}Q_iU_l$ - 6 times 3 dimensions. 

2 generations with 2 $Q$s each can only be made from 2-1 plus 0-1 - but which one is in majority in $QQQ$? In fact it turns out that $(QQQ)_{iij\textbf{h}}Q_j+QQQ_{jji\textbf{h}}Q_i=0$ and $(QQQQ)_{iijj\textbf{d}}\equiv (QQQ_{iij\textbf{h}}Q_j+QQQ_{jji\textbf{h}}Q_i)/\sqrt{2}$ is free. So this adds 3 times 3 dimensions.
Finally we have the case with 2-1-1 disributed over the generations. Here we have the 6 2-1s combined with the last generation of the $Q$ from $QU$ (0-0-1) and the 3 $QQQ_{123\textbf{i}}$ with the 3 possible generations of the $Q$ from the $QU$. The sum of the 3 is still zero, but we will make the definition of the last 2 depend on which generation has the plurality. Thus, with the $Q$ from $QU$ being generation $i$, we want  $QQQ_{ijk\textbf{d}}=(QQQ_{123\textbf{j}}-QQQ_{123\textbf{k}})/\sqrt{2}$ 
 and $QQQ_{ijk\textbf{f}}=(QQQ_{123\textbf{j}}+QQQ_{123\textbf{k}}-2QQQ_{123\textbf{i}})/\sqrt{6}$.
Now
$QQQ_{ijk\textbf{f}}^\al Q_i^\be \ep_{\al\be}/\sqrt{2}+QQQ_{iij\textbf{h}}^\al Q_k^\be \ep_{\al\be}/2+QQQ_{iik\textbf{h}}^\al Q_j^\be \ep_{\al\be}/2=0$ while 
$QQQQ_{iijk\textbf{7}}^a=QQQ_{iij\textbf{h}}^\al Q_k^{\be,a} \ep_{\al\be}/\sqrt{2}-QQQ_{iik\textbf{h}}^\al Q_j^{\be,a} \ep_{\al\be}/\sqrt{2}$, $QQQQ_{iijk\textbf{8}}^a=QQQ_{ijk\textbf{d}}^\al Q_i^{\be,a} \ep_{\al\be}$ and $QQQQ_{iijk\textbf{9}}^a=-QQQ_{ijk\textbf{f}}^\al Q_i^{\be,a} \ep_{\al\be}/\sqrt{2}+QQQ_{iij\textbf{h}}^\al Q_k^{\be,a} \ep_{\al\be}/2+QQQ_{iik\textbf{h}}^\al Q_j^{\be,a} \ep_{\al\be}/2$ are free.

Thus 2-1-1 is 9 times 3 dimensional and  $QQQQU$ is 18*3= 54 dimensional as mentioned in \cite{Gherghetta:1995dv}.
\begin{eqnarray}
QQQQU_{iiijk}&=QQQ_{iij\textbf{h}}^\al Q_i^{\be a} U_k^{\overline{a}} \ep_{\al \be}\de_{a\overline{a}}&[1\leq i,j,k\leq 3,i\neq j]\\
QQQQU_{iijjk}&=(QQQ_{iij\textbf{h}}^{\al} Q_j^{\be a}-QQQ_{jji\textbf{h}}^\al Q_i^{\be a})/\sqrt{2} U_k^{\overline{a}} \ep_{\al \be}\de_{a\overline{a}}&[1\leq i,j,k\leq 3,j=i+1]\\
QQQQU_{iijk\textbf{e}}&=QQQQ_{iijk\textbf{e}}^a U_l^{\overline{a}}\de_{a\overline{a}}&[1\leq i,j,k,l\leq 3,j=i+1,k=j+1,7\leq e\leq 9]
\end{eqnarray}
\subsection{The ones with 3 uncontracted $Q$s}
 Three $Q$s can also be combined without contracting $SU(2)_L$ indices. In \cite{Gherghetta:1995dv}'s notation (exept for the normalisation $\sqrt{6}$) $(QQQ)_\textbf{4}^{\al \be \ga}=Q_i^{\al a}Q_j^{\be b}Q_k^{\ga c}\ep_{abc}\ep^{ijk}/\sqrt{6}$ where the $\textbf{4}$ denotes that the $QQQ$s transform as a $\textbf{4}$ under $SU(2)_L$. \cite{Gherghetta:1995dv} points out that the antisymmetry in family indices are neccessary to balance the antisymmetry in color indices - one can also show by linear algebra that the only free parameter is this one. The only question left to ask is whether it makes a difference which of 3 $SU(2)_L$ doublet fields are contracted with what color index.  It turns out that (with $A,B,C$ being arbitrary $SU(2)_L$ doublets)  $(QQQ)_\textbf{4}^{\al \be \ga}A^{\al '} B^{\be '} C^{\ga '}\ep _{\al \al '}\ep _{\be \be '}\ep _{\ga \ga '}=(QQQ)_\textbf{4}^{\al \be \ga}B^{\al '} A^{\be '} C^{\ga '}\ep _{\al \al '}\ep _{\be \be '}\ep _{\ga \ga '}=(QQQ)_\textbf{4}^{\al \be \ga}B^{\al '} C^{\be '} A^{\ga '}\ep _{\al \al '}\ep _{\be \be '}\ep _{\ga \ga '}$ - in other words the ordering of the other fields is arbitrary.
\begin{eqnarray}
(QQQ_{\textbf{4}}LLH_u)_{i,j}&=(QQQ)_{\textbf{4}}^{\al \be \ga}L_i^{\al '} L_j^{\be '} H_u^{\ga '}\ep _{\al \al '}\ep _{\be \be '}\ep _{\ga \ga '},&[1\leq i\leq j\leq 3]\\
(QQQ_{\textbf{4}}LH_uH_d)_{i}&=(QQQ)_{\textbf{4}}^{\al \be \ga}L_i^{\al '} H_u^{\be '} H_d^{\ga '}\ep _{\al \al '}\ep _{\be \be '}\ep _{\ga \ga '},&[1\leq i\leq 3]\\
QQQ_{\textbf{4}}H_uH_dH_d&=(QQQ)_{\textbf{4}}^{\al \be \ga}H_u^{\al '} H_d^{\be '} H_d^{\ga '}\ep _{\al \al '}\ep _{\be \be '}\ep _{\ga \ga '},&\\
(QQQ_{\textbf{4}}LLLE)_{i,j,k,l}&=(QQQ)_{\textbf{4}}^{\al \be \ga}L_i^{\al '} L_j^{\be '} L_k^{\ga '}E_l\ep _{\al \al '}\ep _{\be \be '}\ep _{\ga \ga '},&[1\leq i\leq j\leq k\leq 3,1\leq l\leq 3]\\
(QQQ_{\textbf{4}}LLH_dE)_{i,j,k}&=(QQQ)_{\textbf{4}}^{\al \be \ga}L_i^{\al '} L_j^{\be '} H_d^{\ga '}E_k\ep _{\al \al '}\ep _{\be \be '}\ep _{\ga \ga '},&[1\leq i\leq j\leq 3,1\leq k\leq 3]\\
(QQQ_{\textbf{4}}LH_dH_dE)_{i,j}&=(QQQ)_{\textbf{4}}^{\al \be \ga}L_i^{\al '} H_d^{\be '} H_d^{\ga '}E_j\ep _{\al \al '}\ep _{\be \be '}\ep _{\ga \ga '},&[1\leq i,j\leq 3]\\
(QQQ_{\textbf{4}}H_dH_dH_dE)_{i}&=(QQQ)_{\textbf{4}}^{\al \be \ga}H_d^{\al '} H_d^{\be '} H_d^{\ga '}E_i\ep _{\al \al '}\ep _{\be \be '}\ep _{\ga \ga '},&[1\leq i\leq 3]
\end{eqnarray}

\subsection{The tricky ones}
Two monomials contain the same superfields as combinations of two other monomials. $UUDQH_uD$ (where the first 3 fields are color contracted) have the same superfields as $UDD*QH_uU$. Clearly, if the 2 $U$s are of the same generation, this new monomial vanishes. If the $D$s are of the same generation, only the new monomial contributes, while the product vanishes. When both $U$ and $D$ have different generations both contribute. On their own, both are two dimentional ($UUD_iQH_uD_j \not \propto UUD_j QH_uD_i$ and $U_iDD*QH_uU_j \not \propto U_jDD*QH_uU_i$) - but there are only three independent combinations of the four - $U_kU_lD_jQH_uD_i=U_kU_lD_iQH_uD_j-U_kD_iD_j*QH_uU_l+U_lD_iD_j*QH_uU_k$ (j=i+1,l=k+1).

Finally, we have $UUDQDQD$ sharing superfields with $UDD*QUQD$. In itself, all combinations of indices (with the two $U$s of different generations) in the new monomials give independent products - 324 in all. Clearly, if all $D$s are of the same generation only the new monomial contributes. By inspection, if the $Q$s are the same, the new monomial does not contribute with extra degrees of freedom\footnote{but they will in general change the basis vectors if one stays in the scheme of making basisvectors orthogonal to what spans zero.}. If $Q$s are different, $U$s are different and one generation of $D$ appears twice and another once, the 4 dimensions of the product (which $U$ in which product - which $Q$ contracts with $U$) and the 3 dimensions of the monomial (where is the ``single'' $D$ placed), are individually independent, but the monomial adds only one degree of freedom - so we just add one of them to the catalogue. 

If there are no superfield generation occuring more than once, the 12 dimensions of the product ($Q$ and $U$ as before - but now three choices for which $D$ is in $QUQD$) can be reduced to 10 - and the monomial adds just a single one of its 6 dimensions (all permutations of $D$ placement). 

All the redundencies of $UUDQDQD$ is shown in the appendix.

\begin{eqnarray}
(UUDQH_uD)_{i,j,k,j}&=U_{i+1}^{\overline{a}}U_{i+2}^{\overline{b}}D_{j}^{\overline{c}}Q_k^{\al a'}H_u^{\be}D_j^{\overline{a'}}\ep_{\overline{a}\overline{b}\overline{c}}\ep_{\al \be}\de_{a' \overline{a'}},&[1\leq i,j,k\leq 3]\\
(UUDQH_uD)_{i,j,k,l}&=U_{i+1}^{\overline{a}}U_{i+2}^{\overline{b}}D_{j}^{\overline{c}}Q_k^{\al. a'}H_u^{\be}D_l^{\overline{a'}}\ep_{\overline{a}\overline{b}\overline{c}}\ep_{\al \be}\de_{a' \overline{a'}},&[1\leq i,j,k,l\leq 3,l=j+1]\\
(UUDQDQD)_{i,j,k+1,j,k+2,j}&=U_{i+1}^{\overline{a}}U_{i+2}^{\overline{b}}D_{j}^{\overline{c}}Q_{k+1}^{\al a'}
D_{j}^{\overline{a'}}Q_{k+2}^{\be b'}D_{j}^{\overline{b'}}\ep_{\al \be}\de_{a' \overline{a'}}\de_{b' \overline{b'}},&[1\leq i,j,k\leq 3]\\
(UUDQDQD)_{i,j,k+1,j,k+2,l}&=U_{i+1}^{\overline{a}}U_{i+2}^{\overline{b}}D_{j}^{\overline{c}}Q_{k+1}^{\al a'}
D_{j}^{\overline{a'}}Q_{k+2}^{\be b'}D_{l}^{\overline{b'}}\ep_{\al \be}\de_{a' \overline{a'}}\de_{b' \overline{b'}},&[1\leq i,j,k,l\leq 3, l\neq j]\\
(UUDQDQD)_{i,1,j+1,2,j+2,3}&=U_{i+1}^{\overline{a}}U_{i+2}^{\overline{b}}D_{1}^{\overline{c}}Q_{j+1}^{\al a'}
D_{2}^{\overline{a'}}Q_{j+2}^{\be b'}D_{3}^{\overline{b'}}\ep_{\al \be}\de_{a' \overline{a'}}\de_{b' \overline{b'}},&[1\leq i,j\leq 3]
\end{eqnarray}

\subsection{Summary of monomials}
Below are the monomials needed to built any gauge invariant operator in  the MSSM and $\nu$MSSM including their dimentionality\footnote{in the tricky cases - only the dimentionality of what is needed to complete the basis} and R-parity. 
\bea
N & 3 & R_- \textbf{ }\nu MSSM \textbf{ only}\\
H_uH_d & 1 & R_+\\
LH_u & 3 & R_-\\
LH_dE & 9 & R_+\\
QH_dD & 9 & R_+\\
QH_uU & 9 & R_+\\
QLD & 27 & R_-\\
LLE & 9 & R_-\\
UDD & 9 & R_-\\
UUDE & 27 & R_+\\
QULE & 81 & R_+ \\
QUQD & 81 & R_+ \\
QQQL & 24 & R_+\\
QUH_dE & 27 & R_-\\
QQQH_d & 8 & R_- \\
DDDLH_d& 3 & R_+ \\
QQQQU  & 54 & R_- \\
QUQUE  & 108 & R_- \\
UUUEE  & 6 & R_- \\
DDDLL  & 3 & R_- \\
UUDQDH_u & 54 & R_- \\
(QQQ)_4LLH_u & 6 & R_- \\
(QQQ)_4LH_uH_d & 3 & R_+ \\
(QQQ)_4H_uH_dH_d & 1 & R_- \\
(QQQ)_4LLLE & 30 & R_- \\
(QQQ)_4LLH_dE & 18 & R_+ \\
(QQQ)_4LH_dH_dE & 9 & R_- \\
(QQQ)_4H_dH_dH_dE & 3 & R_+\\
UUDQDQD & 90 & R_-\\
\textbf{Monomials in total} & 712 (MSSM) & 715 (\nu MSSM)
\eea

\section{Summary and conclusion}\label{scon}
We have presented a complete and minimal catalogue over the 712 monomials from which any gauge invariant operator can be built in the MSSM and, adding three righthanded neutrinos, in the $\nu$MSSM. We have noted that if R-parity is enforced but otherwise all possible gauge invariant couplings are present, in the $\nu$MSSM all flat directions are lifted by terms in the 6th order of the superpotential - compared to order 9 which is needed in the MSSM.
\section{Acknowledgements}
I would like to thank the Danish taxpayers who have supported this work through The Danish Research Council (Forskningsr\aa det) and I thank Stephan Huber, Mark Hindmarsh and David Bailin for useful discussions.

\section{Appendix}

$UUDQDQD$ have no redundencies when the $D$s are all the same - and is only nonzero when the $U$s are different. 
This leaves 4 cases: $D$:2-1 and $D$:1-1-1 combined with different $Q$s  and identical $Q$s. Obviously, it doesn't matter which generations are involved. Therefore the redundencies are only shown for one choice of which generations are involved.

\subsection{$D$:2-1, $Q$s identical}
 Here the monomial, as mentioned in the text, gives no new degrees of freedom.  $D_2U_1U_2Q_1D_2Q_1D_3=U_1D_2D_3*Q_1U_2Q_1D_2-U_2D_2D_3*Q_1U_1Q_1D_2$ is the redundency.

\subsection{$D$:2-1, $Q$s different}
Here the monomial gives one extra degree of freedom.  $D_2U_1U_2Q_1D_3Q_2D_2=-U_1D_2D_3*Q_1U_2Q_2D_2-U_1D_2D_3*Q_2U_2Q_1D_2+U_2D_2D_3*Q_1U_1Q_2D_2+
U_2D_2D_3*Q_2U_1Q_1D_2+D_3U_1U_2Q_1D_2Q_1D_2$ and $D_3U_1U_2Q_1D_2Q_2D_2=-U_1D_2D_3*Q_2U_2Q_1D_2+
U_2D_2D_3*Q_2U_1Q_1D_2+D_3U_1U_2Q_1D_2Q_1D_2$ are the redundencies (other choices possible, since the product span 5 rather than 6 dimensions).

\subsection{$D$:1-1-1, $Q$s identical}
Here the monomial gives no extra degree of freedom.  $D_1U_1U_2Q_1D_2Q_1D_3=U_1D_2D_3*Q_1U_2Q_1D_1+U_2D_3D_1*Q_1U_1Q_1D_2+U_2D_1D_2*Q_1U_1Q_1D_3$, $D_2U_1U_2Q_1D_3Q_1D_1=U_1D_3D_1*Q_1U_2Q_1D_2+U_2D_2D_3*Q_1U_1Q_1D_1+U_2D_1D_2*Q_1U_1Q_1D_3$ and
$D_3U_1U_2Q_1D_1Q_1D_2=U_1D_1D_2*Q_1U_2Q_1D_3+U_2D_2D_3*Q_1U_1Q_1D_1+U_2D_3D_1*Q_1U_1Q_1D_2$ are the redundencies.

\subsection{$D$:1-1-1, $Q$s different}
Here the monomial gives one extra degree of freedom. Here we use a more compact notation:
$P_l=U_iD_{j+1}D_{j+2}*Q_kU_{3-i}Q_{3-k}D_j$ where $1\leq i\leq 2$, $1\leq j\leq 3$, $1\leq k\leq 2$, $l=(i-1)*6+(j-1)*2+k$ $(1\leq l \leq 12)$  and $M_{i,j}=D_iU_1U_2Q_1D_{i+j}Q_2D_{i-j}$ where $1\leq i\leq 3, 1\leq j\leq 2$ and then the redunencies are (other choices possible, since the product span 10 rather than 12 dimensions).
\begin{eqnarray}
 M_{12}&=&P_3+P_4+P_5+P_6+P_7+P_8+M_{11}\\
M_{21}&=&-P_1+P_4+P_7-P_{10}+M_{11}\\
M_{22}&=&P_5+P_{10}-P_{11}+M_{11}\\
M_{31}&=&-P_1-P_2-P_3+P_{10}+P_{11}+P_{12}+M_{11}\\
M_{32}&=&P_4-P_{10}+M_{11}
\end{eqnarray}

\end{document}